\documentclass[prl,superscriptaddress,twocolumn]{revtex4}%
\usepackage[ansinew]{inputenc}
\usepackage{graphicx}
\usepackage{amsmath}
\usepackage{bm}
\usepackage{layout}
\usepackage{float}
\usepackage{txfonts}
\usepackage{amsfonts}
\usepackage{amssymb}%
\begin{document}

\title{Condition for macroscopic realism beyond the Leggett-Garg inequalities}

\begin{abstract}
In 1985, Leggett and Garg put forward the concept of macroscopic realism
(macrorealism) and, in analogy to Bell's theorem, derived a necessary
condition in terms of inequalities, which are now known as the Leggett-Garg
inequalities. In this paper, we discuss another necessary condition called
\emph{no-signaling in time}. It solely bases on comparing the probability
distribution for a macrovariable at some time for the cases where previously a
measurement has or has not been performed. Although the concept is analogous
to the no-signaling condition in the case of Bell tests, it can be violated
according to quantum mechanical predictions even in situations where no
violation of Leggett-Garg inequalities is possible.

\end{abstract}
\date{\today}%

\author{Johannes Kofler}%
%

\affiliation
{Max Planck Institute of Quantum Optics (MPQ), Hans-Kopfermann-Straße 1, 85748 Garching/Munich, Germany}%
%

\author{{\v C}aslav Brukner}%
%

\affiliation
{Faculty of Physics, University of Vienna, Boltzmanngasse 5, 1090 Vienna, Austria}%
%

\affiliation
{Institute for Quantum Optics and Quantum Information, Austrian Academy of Sciences, Boltzmanngasse 3, 1090 Vienna, Austria}%
%

\maketitle

Bell's theorem for local realism~\cite{Bell1964} is a highly developed
research field, not least because of its importance for quantum information
technologies~\cite{Niel2000}. Macroscopic realism
(macrorealism)~\cite{Legg1985}---the world view in which the properties of
macroscopic objects exist independent of and are not influenced by
measurement---has gained momentum only within the past few years as
experiments steadily approached the parameter regime where experimental tests
might become possible. Promising candidates in the race towards an
experimental violation of macrorealism are large superconducting
devices~\cite{Frie2000,Pala2010}, heavy molecules~\cite{Arnd1999,Gerl2011},
and quantum-optical systems in combination with atomic gases~\cite{Buls2001}
or massive objects~\cite{Rome2011}. Still lacking a decisive experiment,
however, the physics community remains to be split into two groups: adherents
of the viewpoint that macrorealism will eventually be falsified by the
preparation of Schrödinger cat-like states~\cite{Schr1935}, and adherents of
one of the hypothetical alternatives saving a classical world on the
macroscopic level~\cite{Ghir1986,Dios1989,Penr1998}.

Macrorealism is defined by the following postulates~\cite{Legg2002}: "(1)
\textit{Macrorealism per se}.\ A macroscopic object which has available to it
two or more macroscopically distinct states is at any given time in a definite
one of those states. (2) \textit{Non-invasive measurability}.\ It is possible
in principle to determine which of these states the system is in without any
effect on the state itself or on the subsequent system dynamics. (3)
\textit{Induction}.\ The properties of ensembles are determined exclusively by
initial conditions (and in particular not by final conditions)."

Since an observation of quantum interference between macroscopically distinct
states (QIMDS), as predicted by quantum mechanics (QM), does not necessarily
establish the falsity of macrorealism, three stages of experiments should be
distinguished~\cite{Legg2002}: "\textit{Stage 1}.\ One conducts circumstantial
tests to check whether the relevant macroscopic variable appears to be obeying
the prescriptions of QM. \textit{Stage 2}.\ One looks for direct evidence for
QIMDS, in contexts where it does not (necessarily) exclude macrorealism.
\textit{Stage 3}.\ One conducts an experiment which is explicitly designed so
that if the results specified by QM are observed, macrorealism is thereby
excluded." Leggett and Garg have put forward the structure of such a stage 3
experiment~\cite{Legg1985}. It consists of measuring temporal correlation
functions and violation of the so called \textit{Leggett-Garg inequalities}.

In this work, we derive a necessary mathematical condition for macrorealism
alternative to the Leggett-Garg inequalities, which we call
\textit{no-signaling in time}. A similar version of this condition was already
discussed in Refs.~\cite{Clif1990,Fost1991,Bena1994} and independently found
in Ref.~\cite{Kofl2008} in the context of coarse-grained measurements of large
spin systems. In Ref.~\cite{Frit2010} the strength of signaling in a temporal
Clauser-Horne-Shimony-Holt scenario was discussed. However, in none of these
references was it recognized as being experimentally more applicable than the
Leggett-Garg inequalities. The condition bases solely on the time evolution of
the probability distribution associated with a macroscopic quantity and can be
viewed as a statistical version of non-invasive measurability. Only two
measurement times are required, while any Leggett-Garg inequality necessarily
involves at least three of them, making a conclusive test of macrorealism more
feasible. We will apply the no-signaling in time condition to the specific
case of interferometric experiments. Once quantum interference between
macroscopically distinct states is shown, it suffices to demonstrate that it
disappears when a prior measurement is made. Our work thus suggests that the
step from a stage 2 to a stage 3 experiment usually can be done in a
straightforward way. We start our analysis with a comparison between local
realism and macrorealism.

Two parties, Alice and Bob, perform measurements on distant particles. Alice's
(Bob's) setting choices are labeled with $a=a_{1},a_{2},...$ ($b=b_{1}%
,b_{2},...$), and her (his) outcomes for a given setting are denoted by $A$
($B$). The assumptions for local realism can be formulated as follows:
\textit{Realism} is a worldview "according to which external reality is
assumed to exist and have definite properties, whether or not they are
observed by someone"~\cite{Clau1978}. \textit{Locality} demands that "if two
measurements are made at places remote from one another the [setting of one
measurement device] does not influence the result obtained with the
other"~\cite{Bell1964}. There is also a third assumption, namely the
\textit{freedom} of choosing the settings independently of the particle
properties. The joint assumption is denoted as local realism (LR) and demands
that the probability for obtaining outcomes $A$ and $B$ under settings $a$ and
$b$ can be written as a convex combination of products of probabilities which
depend only on the local setting and a shared (hidden) variable $\lambda$,
which specifies the properties of every individual particle pair and is
generated with some probability distribution $\rho(\lambda)$%
~\footnote{Equivalently to eq.\ (\ref{eq LR}), local realism guarantees the
existence of a joint probability distribution $P(A_{a_{1}},A_{a_{2}%
},...,B_{b_{1}},B_{b_{2}},...)$ for the results of all possible (mutually
exclusive) measurement results. Here, $A_{a_{1}}$ denotes Alice's outcome for
setting $a_{1}$ and so on. In particular, due to locality, there is no need to
distinguish local outcomes for different distant settings.}:%
\begin{equation}
\text{LR:}\;\;P(A,B|a,b)=\sum\nolimits_{\lambda}\rho(\lambda)\,P(A|a,\lambda
)\,P(B|b,\lambda). \label{eq LR}%
\end{equation}
A special case of local realism is local determinism, where the outcome
probabilities $P(A|a,\lambda)$ and $P(B|b,\lambda)$ are always either 0 or 1.

Quantum mechanics, on the other hand, defines measurement operators $\hat
{M}_{A}^{a}$ and $\hat{M}_{B}^{b}$ for Alice's and Bob's outcomes $A$ and $B$
under settings $a$ and $b$, respectively. The outcome probability for a given
(bipartite) quantum state $\hat{\rho}$ is%
\begin{equation}
\text{QM:}\;\;P(A,B|a,b)=\text{Tr}[\hat{\rho}\,\hat{M}_{A}^{a}\otimes\hat
{M}_{B}^{b}].
\end{equation}

In a basic scenario, there are only two setting choices for each party,
$a=a_{1},a_{2}$, $b=b_{1},b_{2}$, and dichotomic outcomes, $A_{a}=\pm1$,
$B_{b}=\pm1$. According to Bell's theorem, local realism puts a bound on
certain combinations of correlation functions $C_{ab}=\langle A_{a}%
\,B_{b}\rangle$ for distant measurements. This leads to \textit{Bell
inequalities} (BI), e.g.\ the version by Clauser, Horne, Shimony and Holt
(CHSH)~\cite{Clau1969}:%
\begin{equation}
\text{BI:}\;\;C_{a_{1}b_{1}}+C_{a_{2}b_{1}}+C_{a_{2}b_{2}}-C_{a_{1}b_{2}}%
\leq2. \label{eq CHSH}%
\end{equation}
Bell inequalities can be violated by entangled quantum states.

All theories in accordance with the principle of \textit{no-signaling} (NS)
have to ensure that the outcome probabilities for one party must not depend on
the setting of the other party in case the relevant events are space-like
separated:%
\begin{equation}
\text{NS:}\;\;P(B|b)=P(B|a,b)=\sum\nolimits_{A}P(A,B|a,b), \label{eq NS}%
\end{equation}
and vice versa for $P(A|a)$. Here, $\sum\nolimits_{A}P(A,B|a,b)=\sum
\nolimits_{A}P(A|a,b)\,P(B|A,a,b)$, and the sum is taken over all possible
results $A$~\footnote{For uncountable outcomes, integrals instead of sums are
used.}.

Local realism implies both the Bell inequalities and the no-signaling
condition. However, while all local realistic correlations are no-signaling,
the opposite does not necessarily hold. For instance, quantum mechanical
correlations or PR boxes~\cite{Pope1994} are no-signaling but violate local
realism, which means that the correlations cannot be decomposed as in
eq.~(\ref{eq LR}). Thus, no-signaling does not allow to derive Bell's
inequalities. As one cannot reasonably hope to disproove local realism by
observing a violation of the no-signaling condition, it is indeed necessary to
check multiple correlation functions and violate Bell's inequality.

Now we turn to macrorealism. We consider a macroscopic object which is
described by a set of macrovariables $\{Q,Q^{\prime},...\}$, whose values are
considered to be macroscopically distinct by some measure~\cite{Legg2002}.
Examples are the coarse-grained position and momentum for heavy particles in
phase space, the charge and trapped flux in a superconducting quantum
interference device (SQUID), or the magnetic moment along different directions
of large biomolecules. In a series of runs, the object is prepared in the same
initial state, and each preparation defines a new origin of the time axis
$t=0$. Let us consider the case where macrovariable $A\in\{Q,Q^{\prime},...\}$
is measured at time $t_{A}$ ($t_{A}>0$) and macrovariable $B\in\{Q,Q^{\prime
},...\}$ at later time $t_{B}$ ($t_{B}>t_{A}$). (One may of course choose
$A=B$ and measure the same observable twice.) The induction postulate is
reflected by the freedom of choosing the measurement times independently of
the properties of the initially prepared objects. In analogy with
eq.~(\ref{eq LR}), macrorealism predicts that the probability for observing
the outcomes $A$ at $t_{A}$ and $B$ at $t_{B}$ can we written as a convex
combination of products of probabilities where the later measurement outcome
does not depend on the earlier measurement~\footnote{Equivalently to
eq.\ (\ref{eq MR}), macrorealism guarantees the existence of a joint
probability distribution $P(Q_{t_{1}},Q_{t_{2}},...,Q_{t_{1}}^{\prime
},Q_{t_{2}}^{\prime},...)$ for all macrovariables at all times. In particular,
due to non-invasive measurability, there is no need to distinguish later
outcomes for different possible actions at earlier times.}:%
\begin{equation}
\text{MR:}\;\;P(A_{t_{A}},B_{t_{B}})=\sum\nolimits_{\lambda}\rho
(\lambda)\,P(A_{t_{A}}|\lambda)\,P(B_{t_{B}}|\lambda). \label{eq MR}%
\end{equation}
There are two possible ways to define $\lambda$. For every preparation, it can
represent a complete catalogue specifying all properties $\{Q,Q^{\prime
},...\}$ of the object either (i) for all times or (ii) only at the initial
time. In case (i), given $\lambda$, the probabilities $P(A_{t_{A}}|\lambda)$
and $P(B_{t_{B}}|\lambda)$ have to be either 0 or 1. This is due to the
postulate that every macrovariable must always have a definite value.
Stochastic time evolutions are taken into account by a non-trivial
distribution $\rho(\lambda)$, allowing different $\lambda$ even for
identically prepared objects. In case (ii), a complete description
$\lambda_{t}$ of the object at later times $t$ is not determined by $\lambda$,
if the time evolution is stochastic even when $\rho(\lambda)$ is non-zero only
for one $\lambda$. The time evolution of $\lambda$ must not be influenced by
the measurements \footnote{Note the two different levels of 'deterministic
versus stochastic': One level asks whether the complete description at a later
time follows deterministically from the one at an earlier time. The other one
asks whether an individual outcome is determined by the complete description
at the time of measurement. A local realist is free to answer both questions
negatively. As a macrorealist, one is also free regarding the first question
but one has to answer the second one positively. (If you limit yourself to
Newtonian physics, you also need to answer the first question positively.)}.

In contrast to macrorealism, quantum mechanics predicts the outcome
probability%
\begin{equation}
\text{QM:}\;\;P(A_{t_{A}},B_{t_{B}})=\text{Tr}[\hat{\rho}(t_{A})\,\hat{M}%
_{A}]\,\text{Tr}[\hat{\rho}_{A_{t_{A}}}\!(t_{B})\,\hat{M}_{B}].
\end{equation}
Here, $\hat{\rho}(t_{A})$ is the state at time $t_{A}$, $\hat{M}_{A}$ and
$\hat{M}_{B}$ are the measurement operators for outcomes $A$ and $B$, and
$\hat{\rho}_{A_{t_{A}}}\!(t_{B})$ is the (reduced) quantum state at time
$t_{B}$ given that at time $t_{A}$ result $A$ was obtained.

In the simplest case, a single macrovariable $Q$ may only obtain two different
values $Q=\pm1$. Macrorealism restricts the allowed temporal correlations
$C_{t_{A}t_{B}}\equiv\langle Q_{t_{A}}\,Q_{t_{B}}\rangle$ for measurements at
$t_{A}$ and $t_{B}$ and implies the \textit{Leggett-Garg inequalities}
(LGI)~\cite{Legg1985}, e.g.\ of the CHSH type ($t_{1}<t_{2}<t_{3}<t_{4}$):%
\begin{equation}
\text{LGI:}\;\;C_{t_{1}t_{2}}+C_{t_{2}t_{3}}+C_{t_{3}t_{4}}-C_{t_{1}t_{4}}%
\leq2. \label{eq LGI}%
\end{equation}
There is a one-to-one correspondence with the CHSH version of Bell's
inequality~(\ref{eq CHSH}). Alice's and Bob's setting choices correspond to
the measuring times of $Q$ in the following way: $a_{1}\leftrightarrow t_{1}$,
$b_{1}\leftrightarrow t_{2}$, $a_{2}\leftrightarrow t_{3}$, $b_{2}%
\leftrightarrow t_{4}$. (For a discussion about "entanglement in time" see
Ref.~\cite{Bruk2004}.)

A violation of the Leggett-Garg inequality is ubiquitous in the microscopic
quantum
world~\cite{Kofl2007,Kofl2008,Dres2011,Gogg2011,Fedr2011,Wald2011,Knee2012}.
An experimental demonstration of macrorealism with its reference to
macroscopically distinct states is, as pointed out above, still missing. This
is due to the fact that one needs to engineer time evolutions (Hamiltonians)
which build up macroscopic superpositions in time~\cite{Kofl2008} and to
perform multiple temporal correlation measurements before the superpositions
are destroyed by decoherence. Whenever one talks about macroscopic or
classical measurements of quantum systems, one should have in mind
\textit{coarse-grained measurements} which bunch together those quantum levels
to "reasonable" \cite{Pere1995} observables that are neighboring in the sense
of classical physics.

Based on Refs.~\cite{Clif1990,Fost1991,Bena1994,Kofl2008,Frit2010} we now make
the following definition: "\textit{No-signaling in time:} A measurement does
not change the outcome statistics of a later measurement." No-signaling in
time (NSIT) is obeyed by all macrorealistic theories and demands that the
probability for macrovariable $B$ at time $t_{B}$ without any earlier
measurement, $P(B_{t_{B}})$, must be the same as $P(B_{t_{B}|t_{A}})$ where
also an earlier measurement of an arbitrary macrovariable $A$ has been made at
$t_{A}$:%
\begin{equation}
\text{NSIT:}\;\;P(B_{t_{B}})=P(B_{t_{B}|t_{A}})=\sum\nolimits_{A}P(A_{t_{A}%
},B_{t_{B}}). \label{eq SNIM}%
\end{equation}
Here, $\sum\nolimits_{A}P(A_{t_{A}},B_{t_{B}})=\sum\nolimits_{A}P(A_{t_{A}%
})\,P(B_{t_{B}}|A_{t_{A}})$, $P(A_{t_{A}})$ is the probability for result $A$
at $t_{A}$, and $P(B_{t_{B}}|A_{t_{A}})$ is the probability for outcome $B$ at
$t_{B}$, given result $A$ was obtained at $t_{A}$. If we denote the
probability amplitudes for the results $A$ by $a_{A}$ ($|a_{A}|^{2}%
=P(A_{t_{A}})$) and the transition probability amplitudes from $A$ to $B$ by
$a_{A\rightarrow B}$ ($|a_{A\rightarrow B}|^{2}=P(B_{t_{B}}|A_{t_{A}})$), then
the difference between the left and right hand side of~(\ref{eq SNIM}) reads
$P(B_{t_{B}})-P(B_{t_{B}|t_{A}})=|\!\sum\nolimits_{A}a_{A}\,a_{A\rightarrow
B}|^{2}-\sum\nolimits_{A}P(A_{t_{A}})\,P(B_{t_{B}}|A_{t_{A}})$. This shows
that the violation of no-signaling in time~\footnote{The overlap between the
two measured probability distributions is, for example, quantified by
$\kappa\equiv%
{\textstyle\sum\nolimits_{B}}
\!\sqrt{P(B_{t_{B}})P(B_{t_{B}|t_{A}})}\in\lbrack0,1]$. A violation of
(\ref{eq SNIM}) can be stated if identical distributions ($\kappa=1$) are
ruled out in a statistically significant way. The issues of finite data sets,
measurement inaccuracies, and statistical significance also arise in tests of
Bell or Leggett-Garg inequalities.} is exactly given by the quantum mechanical
interference terms~\cite{Daki2012}.

No-signaling in time is the analog of the no-signaling condition~(\ref{eq NS})
in the sense that both are operationally testable and can be viewed as
statistical versions of non-invasive measurability and locality, respectively.
The key difference is that a violation of no-signaling in time is not at
variance with special relativity and can be achieved by quantum mechanics.

Macrorealism implies both the Leggett-Garg-inequalities and no-signaling in
time. But the latter does in general not allow to derive Leggett-Garg
inequalities, which can be seen from the following (thought) experiment:
Consider an ensemble, where initially, at time $t=0$, half of the systems are
in the state $Q=+1$ and the other half in $Q=-1$. Let the time evolution be
the macroscopic analog of a precessing spin-$\frac{1}{2}$ particle with
frequency $\omega$~\cite{Kofl2007}. Macroscopic quantum superpositions are
produced in time in such a way that the temporal correlation function for
measurements at times $t_{A}$ and $t_{B}$ reads $C_{t_{A}t_{B}}\equiv\langle
Q_{t_{A}}\,Q_{t_{B}}\rangle=\cos[\omega(t_{B}\!-\!t_{A})]$. Suitable
measurement times allow for a (maximal) violation of the Leggett-Garg
inequality~(\ref{eq LGI}), while no-signaling in time is still fulfilled
between any pair of measurements: $P(Q_{t_{B}}\!=\!+1)=P(Q_{t_{B}|t_{A}%
}\!=\!+1)=\frac{1}{2}$. Due to the mixedness of the initial state, the
violation of macrorealism can hide in the statistics of
condition~(\ref{eq SNIM}). However, if a given Hamiltonian permits to violate
macrorealism, then any initial pure state allows one to find time instances
$t_{A}$ and $t_{B}$ such that no-signaling in time can be violated. To show
this, it is enough to notice that a violation of macrorealism requires
interference of superposition branches. An intermediate measurement destroys
the interference term and thus makes itself detectable at a later
time.\begingroup\begin{table}[t]
\begin{ruledtabular}
\begin{tabular}
[c]{p{3cm}p{4.5cm}}%
Local Realism (LR) & Macrorealism (MR)\\[3pt] \hline
Bell inequality (BI) & Leggett-Garg inequality (LGI)\\
No-signaling (NS) & No-signaling in time (NSIT)\\
LR $\Rightarrow$ BI & MR $\Rightarrow$ LGI\\
LR $\Rightarrow$ NS & MR $\Rightarrow$ NSIT\\
NS $\nRightarrow$ BI & NSIT $\nRightarrow$ LGI\\
QM $\nRightarrow$ BI & QM $\nRightarrow$ LGI\\
\emph{QM} $\Rightarrow$ \emph{NS} & \emph{QM} $\nRightarrow$
\emph{NSIT}
\end{tabular}
\end{ruledtabular}
\caption{Local realism and macrorealism are largely analogous in their
conceptual relationships. The Bell and Leggett-Garg inequalities are both
violated by quantum mechanics (QM). The key difference (written in italics) is
that quantum mechanics obeys no-signaling while it can violate no-signaling in
time.}%
\label{tab}%
\end{table}\endgroup

Table~\ref{tab} sums up the conceptual relationships in local realism and
macrorealism. It is worth mentioning that in principle there exist situations
where no-signaling is violated although no sufficient number of setting is
involved to construct a Bell inequality. This is the case when Bob has only
one possible setting, say $b_{2}$, and his outcome $B$ reveals Alice's setting
choice. Such a model violates local realism~(\ref{eq LR}) and
no-signaling~(\ref{eq NS}), but no Bell inequality can be constructed with
only two correlation functions $C_{a_{1}b_{2}}$ and $C_{a_{2}b_{2}}$.
Similarly, if only two temporal correlations $C_{t_{1}t_{4}}$ and
$C_{t_{3}t_{4}}$ are allowed to be measured, no Leggett-Garg inequality can be
violated. However, a violation of macrorealism~(\ref{eq MR}) and no-signaling
in time~(\ref{eq SNIM}) remains detectable.

To exemplify the usefulness of no-signaling in time, we consider a double slit
experiment with large objects. Each object is emitted at the time $t_{0}$,
passes a double slit at time $t_{1}$, and arrives at a detection screen at
time $t_{2}$. As macrovariable we choose the lateral position variable denoted
by $x$. We assume that the slit distance $d$ is large enough to qualify for
the term "macroscopically distinct". Now three experiments are performed, each
with many runs: I.~Both slits are open. II.~The left slit is blocked by a
detector. Only objects passing the right slit will reach the detection screen
at $t_{2}$. These are the ones which, according to a macrorealist, cannot be
influenced by the measurement at $t_{1}$ at the other slit (ideal negative
result measurements~\cite{Legg1985,Legg1988}). III.~Same as experiment II but
with the right slit blocked. No-signaling in time predicts that the
distribution $P_{\text{I}}(x_{t_{2}})$ found in experiment I must be the same
as the weighted mixture $P_{\text{II\&III}}(x_{t_{2}|x_{t_{1}}})$ of the
single-slit distributions found in experiments II and III. Quantum mechanics,
on the other hand, predicts interference between macroscopically distinct
states, and thus an interference pattern for $P_{\text{I}}(x_{t_{2}})$, but no
interference fringes for $P_{\text{II\&III}}(x_{t_{2}|x_{t_{1}}})$. It is
important to note that there seems to be no way to write down a violable
Leggett-Garg inequality for the double-slit experiment.

For a Mach-Zehnder interferometer, a three-time Leggett-Garg inequality of the
Wigner form~\cite{Wign1970} $C_{t_{0}t_{1}}+C_{t_{1}t_{2}}-C_{t_{0}t_{2}}%
\leq1$ can be employed. Before the first beam splitter (time $t_{0}$), inside
the interferometer (time $t_{1}$), and after the second beam splitter (time
$t_{2}$), there are always two possible paths ($Q_{t_{0}},Q_{t_{1}},Q_{t_{2}%
}\!=\!\pm1$). Depending on the four parameters---the initial probability
distribution for $Q_{t_{0}}$, two reflectivities, and one phase shift in the
interferometer---one can find regimes where the Leggett-Garg inequality is
obeyed but, according to quantum mechanics, no-signaling in time is violated
and others where the opposite is the case. For instance, if both beam
splitters are balanced with reflectivities $\frac{1}{2}$, the Leggett-Garg
inequality becomes $-C_{t_{0}t_{2}}\leq1$ and cannot be violated. No-signaling
in time, on the other hand, demands $P(Q_{t_{2}})=P(Q_{t_{2}|t_{1}})$ and is
violated by an intermediate measurement inside the interferometer at $t_{1}$
for all parameter choices which allow interference. In contrast, for an
initial mixture $P(Q_{t_{0}}\!=\!\pm1)=\frac{1}{2}$, no-signaling in time
cannot be violated, while the Leggett-Garg inequality can be violated for a
suitable choice of reflectivities and phase. (For more details see the
Appendix.) This demonstrates that neither the violation of the 3-time
Leggett-Garg inequality nor the violation of no-signaling in time is necessary
for a violation of macrorealism.

Neither realism nor macrorealism per se can be tested on their own, which is
why experimental tests have to be carefully designed to avoid loopholes. The
three main loopholes in Bell experiments---locality, fair sampling, and
freedom of setting choice---have all been closed
individually~\cite{Aspe1982,Weih1998,Rowe2001,Ansm2009,Sche2010,Gius2013}. In
macrorealism, the non-invasiveness loophole should be closed by performing
ideal negative result measurements~\cite{Legg1985,Legg1988}, which has been
achieved already for microscopic systems~\cite{Knee2012}. Closing the fair
sampling and freedom-of-choice loopholes will require high detection
efficiency and statistical independence between the measured macroscopic
object and the chosen measurement times just as in Bell tests.

Assume that one day a loophole-free experiment is performed which violates the
Leggett-Garg inequality (or no-signaling in time) for macroscopic observables,
thereby ruling out objective collapse
theories~\cite{Ghir1986,Dios1989,Penr1998}. The Bohmian interpretation of
quantum mechanics~\cite{Bohm1952} would still claim a well-defined position
for every object at all times and allow for a description obeying macrorealism
per se (as it would still provide a realistic description of a loophole-free
Bell test). Ideal negative result measurements do not change the position
macrovariables themselves, but they alter their subsequent time evolution due
to an instant (non-local) change of the quantum wave function which serves as
a guiding potential, thus violating the non-invasiveness condition. Bohmian
mechanics is realistic, non-local, and no-signaling in the language of Bell,
and it is macrorealistic per se, invasive, and signaling in time in the
language of Leggett-Garg and the present work. Even if, as the authors, one
does not adhere to this interpretation, this indicates a deeper connection
between locality and non-invasive measurability beyond their formal analogy.

\textit{Conclusion.} We have identified "no-signaling in time" as an
alternative necessary condition for macrorealism which is different from the
Leggett-Garg inequalities. Both conditions, no-signaling in time and the
Leggett-Garg inequalities, are implied by macrorealism, but in general neither
implies the other and neither violation is necessary for a violation of
macrorealism. However, there are two main advantages of no-signaling in time,
making it appealing for future experiments: 1.\ While a Leggett-Garg test
needs to involve at least three possible measurement times, no-signaling in
time requires only two, allowing for tests in situations, where Leggett-Garg
inequalities cannot be used at all. 2.\ As no-signaling in time can be
violated by any non-vanishing interference term, it usually can be violated
for a much wider parameter regime than the Leggett-Garg inequalities. Finally,
one might argue that a violation of no-signaling in time is a direct violation
of non-invasive measurability and that one is interested only in those
situations where no-signaling in time is obeyed but a Leggett-Garg inequality
is violated. However, we note that a violation of no-signaling in time---just
as a violation of the Leggett-Garg inequality---can be achieved using ideal
negative measurements. Violating no-signaling in time is thus no more a
violation of non-invasive measurability (or of macrorealism per se) than a
violation of the Leggett-Garg inequality itself. In summary, our work has
shown that the step from a stage 2 experiment (showing quantum interference of
macroscopically distinct states) to a stage 3 experiment (stage 2 and
simultaneously ruling out macrorealism) does not require the complexity of the
Leggett-Garg inequalities. It suffices to test the simpler criterion of
no-signaling in time.

\textit{Note added in proof}. Recently, a related work~\cite{Li2012} has been
submitted and published.

\textit{Acknowledgments.} We acknowledge discussions with M.\ Aspelmeyer,
J.\ I.\ Cirac, B.\ Daki\'{c}, O.\ Romero-Isart, and A.\ J.\ Leggett.

\subsection{Appendix}

\begin{figure}[t]
\begin{center}
\includegraphics{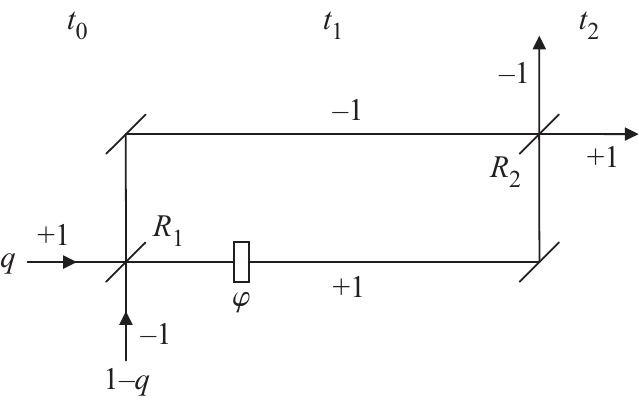}
\end{center}
\par
\vspace{-0.5cm}\caption{Schematic of the Mach-Zehnder interferometer.}%
\label{figure}%
\end{figure}

Let us consider in more detail the example of a Mach-Zehnder interferometer
(see Fig.~\ref{figure}). Because there are always two possible paths before
the first beam splitter (time $t_{0}$), inside the interferometer (time
$t_{1}$), and after the second beam splitter (time $t_{2}$), the macrovariable
$Q$ can always take one of two possible values: $Q_{t_{0}},Q_{t_{1}},Q_{t_{2}%
}\!=\!\pm1$. The reflectivities of the first and second beam splitter are
denoted by $R_{1}$ and $R_{2}$, and the phase shift in the lower arm is called
$\varphi$. We choose an arbitrary initial distribution $P(Q_{t_{0}}%
\!=\!+1)=q$, $P(Q_{t_{0}}\!=\!-1)=1-q$ of incoming objects in a mixed quantum
state $q\left\vert +1\right\rangle \!\left\langle +1\right\vert
+(1-q)\left\vert -1\right\rangle \!\left\langle -1\right\vert $ with
$\left\vert +1\right\rangle $ and $\left\vert -1\right\rangle $ corresponding
to the macrovariable values $+1$ and $-1$, respectively.

Quantum mechanics predicts the following temporal correlations $C_{t_{i}t_{j}%
}=%
{\textstyle\sum\nolimits_{k,l=\pm1}}
\,k\,l\,P(Q_{t_{i}}\!=\!k,Q_{t_{j}}\!=\!l)$ between times $t_{i}$ and $t_{j}$:%
\begin{align}
C_{t_{0}t_{1}}  &  =1-2\,R_{1},\nonumber\\
C_{t_{1}t_{2}}  &  =2\,R_{2}-1,\\
C_{t_{0}t_{2}}  &  =-1+2\,R_{1}+2\,R_{2}-4\,R_{1}R_{2}+4\sqrt{R_{1}T_{1}%
R_{2}T_{2}}\cos\varphi,\nonumber
\end{align}
with the transmittances $T_{1}=1-R_{1}$ and $T_{2}=1-R_{2}$. The correlation
functions are independent of $q$ because the first measurement always acts as
a preparation for the second. The Leggett-Garg inequality $K\equiv
C_{t_{0}t_{1}}+C_{t_{1}t_{2}}-C_{t_{0}t_{2}}\leq1$ reads%
\begin{equation}
1-4\,R_{1}T_{2}-4\!\sqrt{R_{1}T_{1}R_{2}T_{2}}\cos\varphi\leq1.
\end{equation}
The maximum violation of $K=1.5$ is achieved for $R_{1}=\frac{1}{4}$,
$R_{2}=\frac{3}{4}$, and $\varphi=\pi$. No violation is possible, e.g., if
$R_{1}=$ $R_{2}=\frac{1}{2}$, because then $C_{t_{0}t_{1}}$ and $C_{t_{1}%
t_{2}}$ vanish and the remaining inequality $-C_{t_{0}t_{2}}\leq1$ is always fulfilled.

No-signaling in time demands%
\begin{equation}
P(Q_{t_{2}}\!=\!+1)=\sum\nolimits_{Q_{t_{1}}=\pm1}P(Q_{t_{1}},Q_{t_{2}%
}\!=\!+1).
\end{equation}
The left hand side is the probability for outcome $+1$ at time $t_{2}$ without
any prior measurements and is, according to quantum mechanics, given by
$\frac{1}{2}+\frac{1}{2}\,(2q-1)\,C_{02}$. The right hand side is the same
probability but in the case of an intermediate measurement at time $t_{1}$ and
is given by $\frac{1}{2}+(2q-1)\,(-\frac{1}{2}+R_{1}+R_{2}-2\,R_{1}R_{2})$.
The difference between the left and the right hand side, which must vanish if
no-signaling in time holds, reads%
\begin{equation}
2\,(2q-1)\!\sqrt{R_{1}T_{1}R_{2}T_{2}}\cos\varphi=0.
\end{equation}
This is violated whenever the parameters allow for interference, i.e., when
neither of the reflectivities is 0 or 1, the phase is unequal to $\frac{\pi
}{2}$ and $q$ is unequal to $\frac{1}{2}$. The biggest violation (largest
interference) is achieved for $R_{1}=$ $R_{2}=\frac{1}{2}$, $q=0$ or $1$, and
$\varphi=0$ or $\pi$.

For many parameter choices the Leggett-Garg inequality and no-signaling in
time are both violated or both fulfilled. However, there are also parameter
regimes (e.g., $R_{1}=\frac{1}{4}$, $R_{2}=\frac{3}{4}$, $\varphi=\pi$, and
$q=\frac{1}{2}$) for which quantum mechanics violates the Leggett-Garg
inequality while no-signaling in time is fulfilled, and others (e.g.,
$R_{1}=\frac{1}{2}$, $R_{2}=\frac{1}{2}$, $\varphi=\pi$, and $q=1$) for which
no-signaling in time is violated while the Leggett-Garg inequality is
satisfied. This demonstrates in general that---although both the Leggett-Garg
inequality and no-signaling in time are a consequence of
macrorealism---neither of these two criteria implies the other. Therefore,
neither the violation of the Leggett-Garg inequality nor the violation of
no-signaling in time is a necessary condition for a violation of macrorealism.

\end{document}